\documentstyle[12pt]{article}
\begin{document}

\begin{center}

\ \vskip 1.cm  
{\Large
{\bf 
Lateral superlattices as voltage-controlled traps 
for excitons}}

\vskip 0.6cm

S. Zimmermann, 
A. O. Govorov$^{*}$, W. Hansen$^{**}$, and
J. P. Kotthaus 
\vskip 0.4cm

{\it Sektion Physik der Ludwig-Maximilians-Universit\"at,  
Geschwister-Scholl-Platz 1, \\ 
D-80539 M\"unchen, Germany}

\vskip 0.4cm
M. Bichler and W. Wegscheider

\vskip 0.4cm

{\it Walter Schottky Institut, Technische Universit\"at M\"unchen, D-85748 Garching, 
Germany}

\end{center}

\vskip 0.6cm

We demonstrate the localisation of quantum well excitons in a periodic array of linear traps
using photoluminescence experiments. The excitonic traps are induced by applying 
spatially alternating external voltages via interdigitated metal gates. The localisation 
originates from the periodical modulation of the strength of the quantum-confined Stark effect in 
the plane of the quantum well.  In our experiments, the trap depth is easily tuned by 
the voltages applied to the interdigitated gates. Furthermore, we find that a 
perpendicular magnetic field reduces the exciton diffusion length. In short-period lateral 
superlattices, we observe a magnetic-field-induced stabilisation of excitons in the presence of 
strong in-plane electric fields. 

\newpage

{\large Introduction }
\vskip 0.4cm

Electro-optic interactions involving excitons in semiconductor microstructures and microdevices are
currently attracting much interest \cite{Schmitt-Rink}. Such devices are usually laterally 
microstructured in the plane of quantum wells. Consequently, it is interesting to study 
quantum well excitons in laterally varying potentials and, in particular, the
possibility of localising them in distinct quasi-one- or quasi-zero-dimensional 
regions of a sample. Whereas the localisation of charged particles, like electrons or holes, is easily possible in  electrostatic lateral potential superlattices \cite{Drexler94}, 
the localisation of 
neutral, but polarisable, excitons needs to make use of a more special mechanism. The spatial 
localisation of excitons in one- and zero-dimensional structures has been realised in 
semiconductors by various approaches \cite{Kash,Brunner,Marzin}. One mechanism to 
induce exciton localisation is based on strain \cite{Kash}. Other possibilities include inter-diffusion of a barrier material \cite{Brunner} and the preparation of 
self-organised lateral structures \cite{Marzin}. In all these cases, the exciton localisation cannot be changed by externally tuneable parameters. In the present study, we realise a 
voltage-tuneable localisation for excitons within the plane of a quantum well. 
   
Exciton transport in semiconductors can be induced by spatially non-uniform electric fields 
via the Stark effect. The change of the exciton energy by the electric field is $-dE$, where $d$ is the exciton dipole moment induced by the electric field $E$. 
The energy $-dE$ plays the role of 
the effective exciton potential, which will be denoted in the following as $U_{eff}$.
 If the 
electric field is dependent on the spatial coordinate, the potential $U_{eff}$ can form excitonic traps. In principle, the trap potential $U_{eff}$ can be controlled by external voltages. 
At low 
temperatures this mechanism of exciton localisation is effective if the localisation energy is 
significantly larger than the spatially random fluctuations in the exciton energy, 
which result in an inhomogeneous broadening of the exciton peak. 
According to 
numerical calculations \cite{Lederman}, for the case of bulk excitons in GaAs the maximal 
Stark red- shift is less than $1 \ meV$.
In a strictly two-dimensional (2D) system, the calculated value of the Stark red- 
shift caused by an {\it in-plane } electric field does not exceed $2 \ meV$ in GaAs 
\cite{Lederman}.
This limitation is imposed by exciton ionisation in electric fields beyond 
approximately $10^4 \ V/cm$. 
The typical broadening of the exciton peak in GaAs-based systems is about $1 \ meV$. 
Since for the localisation of excitons the Stark shift needs to be larger than the broadening of the exciton peak, 
the Stark effect in bulk systems as well as the in-plane Stark effect in quantum wells 
\cite{exp} can hardly be used to localise excitons. 

To realise effective excitonic traps in quantum wells (QW's), we make use 
of the quantum-confined Stark effect (QCSE) \cite{Miller}. 
The QCSE arises from the electric field applied 
$ perpendicular $ to a QW. For a QW of $16 \ nm$ width, the Stark shift in a perpendicular 
electric field of $10^5 \ V/cm$ is about $50 \ meV$ \cite{Vina}, and thus is sufficiently large for an 
effective localisation of excitons. In such strong perpendicular electric fields, the exciton 
remains stable because of the QW confinement. 
The exciton localisation induced by the QCSE can be effective in 
relatively wide QW's, where the exciton energy shift due to the QCSE exceeds the 
inhomogeneous broadening of the exciton peak. 
A laterally periodic modulation of the effective exciton potential $U_{eff}$ 
can be obtained in a quantum well system with an interdigitated top gate and 
a doped back contact 
 as shown in Fig. 1. This design allows us to induce a strong lateral
modulation of the perpendicular electric field causing the spatially-modulated QCSE.  In our 
scheme shown in Figs. 1 and 2 the spatial modulation of the QCSE occurs when we apply the 
different voltages $V_1$ and $V_2$ to the finger gates with respect to the back contact.  The 
effective exciton potential induced by the interdigitated gate leads to a periodic array of 
excitonic traps in the sample.
   
Initial results on exciton localisation in a QW system with an interdigitated gate of a small 
period have been published in a conference report by A. Schmeller et al. \cite{Schmeller95}. In this study, the photoluminescence signal arose from excitons localised in linear traps via the QCSE as described above. 
More recently, the lateral transport of indirect excitons was studied in a QW 
system by use of two surface metal gates, spatially separated by a macroscopic 
distance \cite{Markus}. 
These surface gates were differently biased. 
Exciton transport in such a system is also based on the QCSE 
in a  varying electric field and was studied with a spatially resolved technique \cite{Markus}.  
   
Here we present a detailed study of the photoluminescence (PL) of a single-QW system with interdigitated metal gates of various periods, also in the presence of a magnetic field. In the samples investigated, the lateral gate periods are comparable to the exciton diffusion length. This enables us to study the exciton diffusion in more detail. 
All the specific features of the PL spectra are found to be consistent with our model of an effective exciton potential. The mechanism of exciton localisation is established to be the QCSE.
Along with the experimental study, we model theoretically the lateral exciton transport in a spatially modulated potential. Using our experimental data, we are able to estimate the exciton diffusion length.
We find that the diffusion length is strongly reduced in the presence of a magnetic field.

\vskip 0.5cm
{\large 1. Experimental details}
\vskip 0.4cm

The AlAs-GaAs heterostructure employed here was grown by molecular beam epitaxy and 
consists of a Si-doped back electrode ($N_D=4*10^{18} \ cm^{-3}$) grown on an undoped 
GaAs buffer, and a $20$-$nm$-QW embedded in a short-period AlAs-GaAs superlattice barrier. 
The QW and the back electrode are separated from the surface by $60 \ nm$ and $390 \ nm$, respectively.
On top of the heterostructure we deposit semi-transparent titanium gates in an interdigitated geometry as shown in Figs. 1 and 2 a. 
The interdigitated gate is characterised by its period $a$ and stripe width $w$. We have fabricated samples with three different gate parameters: 
$(a,w)=(1500 \ nm, \ 1300 \ nm)$, $(1000 \ nm, \ 900 \ nm)$, and $(250 \ nm, \ 190 \ nm)$. 
The finger gates are biased with respect to the back contact by voltages $V_1$ and $V_2$, 
respectively. The resulting electrostatic potential in the sample is laterally modulated with 
period $2a$. When the voltages $V_1$ and $V_2$ are slightly above zero, the QW 
starts to become occupied  
 by electrons. The experiments discussed here are performed with gate 
 voltages at which there is no charge accumulation in the QW.

To study the PL we excite the sample by a diode laser with photon energy $1.58 \ eV$. The 
sample is mounted in a variable temperature insert of 
a magneto-optical cryostat.
The intensity  of the incident light is about $0.5 \ W/cm^2$. 
The PL study is performed at temperature $T=3.5 \ K$, so that the PL is of excitonic origin.

\vskip 0.5cm
{\large 2. General behavior of the photoluminescence spectra}
\vskip 0.4cm

First, we consider the PL spectrum of a QW with a lateral superlattice of period 
$2a=3000 \ nm$. 
According to our numerical calculations for this sample, the electrostatic 
potential in the plane of the QW nicely reproduces the potentials of the finger gates and is close to being rectangular. 
The form of the electrostatic potential is determined by the geometrical parameters 
of the sample where in this case the distance between the surface and the plane 
of the QW is much less than the lateral size of the metal stripes. 
Optical and electronic properties of similar quantum well devices with interdigitated gates were studied before in a series of experiments  \cite{Drexler94,Schmeller95,Schmeller94}.
In particular, it was demonstrated that such a system gives us the possibility to control independently the average potential and the amplitude of the potential modulation in the plane of a QW. 
When the voltages $V_1$ and $V_2$ applied to the corresponding finger gates are equal, 
the potential modulation in the system is nearly absent \cite{remark0}. 
If the voltages $V_1$ and $V_2$ are different, the electrostatic potential 
in the system is modulated to form a lateral superlattice with period $2a$ and the amplitude of the potential modulation is determined by the voltage difference 
$\Delta V=V_1-V_2$.  
   
If $V=V_1=V_2$, the PL spectrum exhibits a single narrow heavy-hole-exciton peak 
(full width at half maximum $FWHM=4.1 \ meV$), which is shifted with voltage $V$ in 
accordance with the QCSE (see  Fig. 3). 
The voltage dispersion of the PL peak is as expected for a $20$-$nm$-QW. 
In fact, at voltages $V_1=V_2$ the interdigitated gate system behaves essentially 
like a QW system with a homogeneous metal gate \cite{Schmeller95}.  Hence, the data for 
$V_1=V_2$ can be used as a reference. 

With application of different voltages $V_{1}$ and $V_{2}$
the PL spectrum is modified as shown in Figs. 4 a, b. 
These spectra are obtained when the average voltage is kept 
constant: $(V_1+V_2)/2=-0.9 \ V$. 
At the same time, the voltage difference $\Delta V=V_1-V_2$ is varied from $0$ 
to $-1.8 \ V$ in steps of $0.1 \ V$.  In the spectra shown in Figs. 4 
a, b, we observe a splitting of the exciton peak into two features.
The splitting increases with increasing voltage difference $ \Delta V$. 
The energy positions of the peaks are shifted in opposite directions. 
In the following, we will call these structures low and high energy peaks (LEP and HEP). The 
energy positions of the LEP and the HEP at zero magnetic field are shown in Fig. 3 by filled symbols as a function of the voltages $V_1$ and $V_2$, respectively. 
In a magnetic field of $7 \ T$, the energies of all peaks are increased approximately by $5 \ meV$ according to the diamagnetic shift.  
   
In our long-period superlattices we may expect that the positions of 
the LEP and the HEP correspond to the local perpendicular electric fields 
beneath the gates induced by the voltages $V_{1}$ and $V_{2}$, respectively.  
The energy positions of the LEP and the exciton peak in a homogeneous system (open 
symbols 
in Fig. 3) coincide. Thus, the energy of the LEP follows the voltage $V_1$. The energy 
positions of the HEP in Fig. 3 are a few meV higher with respect 
to the open symbols related to a 
homogeneous system.
A likely cause of this difference is a redistribution of surface charges with the application of the voltage difference. 
Despite this minor discrepancy for the HEP, the general behavior of the LEP and the HEP 
corresponds to the QCSE induced by the voltages $V_1$ and $V_2$, respectively. 
Thus, we conclude that the LEP and the HEP are connected with the PL signal arising from the regions under the gates 1 and 2, respectively. 
   
Now we would like to discuss the width of the PL peaks in Figs. 4 a, b. 
As an example, consider the 
spectra for magnetic field $7 \ T$.  At high voltage differences, the $FWHM$ of the 
LEP and the HEP are equal to $3.1 \ meV$ and $2.4 \ meV$,  respectively. 
This variation of 
the $FWHM$ can be understood keeping in mind that the LEP and the HEP arise from 
regions of the sample with different perpendicular electric fields.
The width of an exciton peak in the QCSE regime for quantum wells increases with the normal electric field \cite{Schmeller95} 
because of interface-roughness-induced scattering. 
The $FWHM$ of the exciton peak in Fig. 4  at zero voltage difference is $4.1 \ meV$, i. e. larger than those of the LEP and the HEP at high voltage differences. 
We can explain this observation with a residual 
potential modulation in the sample at zero voltage difference, which is induced by a surface potential variation between the metallic grating stripes and free surface stripes in between.

\vskip 0.2cm
\ \\
{\large 3. The PL intensity as a function of the voltage difference and the 
magnetic field}
\vskip 0.4cm

In this section, we consider the dependence of the LEP- and HEP-intensities on the 
voltage difference $\Delta V$.  We start our discussion with the PL spectrum recorded at strong magnetic fields (see Fig. 4 b), because the HEP becomes more pronounced with increasing magnetic field.  
One can see from Fig. 4 b that the HEP becomes visible at voltage difference $0.1 \ V$. Initially, with increasing voltage difference $\Delta V$ the intensity of the LEP is much 
larger than that of the HEP. 
At high voltage differences, the intensities of the LEP and the HEP become comparable. The 
reduction of the HEP intensity at $V_2 \approx 0 \ V$ is likely to be caused by 
the appearance of electrons beneath gate 2. 
  
The above-mentioned behavior of the PL spectra can be explained by a simple model of the 
effective exciton potential as sketched in Fig. 2b. 
In this figure we distinguish between the lateral regions  I, II, and  
III,  that are located beneath the gates 1 and 2, and in between the gates, respectively.
The PL 
spectra shown in Figs. 4 a, b correspond to the voltage regime
in which the exciton energy decreases with decreasing gate voltage. 
The effective exciton potential, $U_{eff}$, for this regime is 
shown qualitatively in Fig. 2b for the case of an applied gate voltage difference $\Delta V$.
In this model, the regions I correspond to exciton traps, 
while 
the regions II are areas of maximum exciton energy.  The incident light creates excitons in all 
regions of the sample. The typical diffusion length of an exciton in QW's is of the order of 
$1 \ \mu m$ \cite{Heller}.  Because the lateral sizes of our samples are comparable to the 
exciton 
diffusion length, most of the excitons are able to reach the potential traps due to diffusion and 
drift.  
Thus, we can expect that regions I collect excitons from all areas of the sample and that the LEP should be dominant in the PL spectrum. 

Indeed, we observe this behavior in the experimental 
spectra at small voltage differences. The HEP occurs because some excitons recombine before 
they reach the potential minima.  Another prominent feature of the experimental spectra is a 
decrease of the 
total PL intensity with increasing voltage difference
$\Delta V$. 
We associate this observation with ionisation of 
excitons by the in-plane electric field.  
Strong in-plane electric fields in the sample are induced by 
the voltage difference $\Delta V$ and 
are maximal in the regions III. 

Figure 5 shows the voltage-dependence of the 
LEP intensity.  One can see that the intensity of the 
LEP decreases to half of its initial value at $V_1=-0.975 \ V$, corresponding to a voltage difference of $0.15 \ V$. 
From numerical calculations of the electrostatic potential in our sample we find that a voltage 
difference of $0.15 \ V$ corresponds approximately to an in-plane electric field in regions 
III of $4\times10^3 \ V/cm$. This is a typical ionisation electric field for an exciton in bulk 
GaAs \cite{Lederman} and in relatively wide quantum wells.  Thus, at voltage 
differences larger 
than $0.15 \ V$ the regions I and II are no longer coupled by exciton transport because all 
excitons moving from the regions II to the regions I are ionised by 
the strong in-plane electric field in regions III. 

It can be seen from Fig. 5 that the LEP intensity decreases slowly at 
 higher voltage 
differences and its magnitude becomes significantly less than the surface fraction of gate 1, 
$1300nm/3000nm=0.43$. We associate this behavior with a reduction of the exciton peak intensity at 
high perpendicular electric fields which is typical for the QCSE 
regime even in homogeneous systems \cite{Vina}.

The spectra of Fig. 4 b show that at a high voltage difference $\Delta V$
and at 
 a magnetic field $B=7 \ T$ the 
asymmetry between the peaks is strongly reduced but the intensity of the LEP is still larger 
than that of the HEP. At high voltage differences the regions III with a strong in-plane electric 
field act as drains, where the excitons are ionised.  In the regions I, the effective exciton 
potential is minimal in the center, and, 
consequently, excitons are localised in the areas of a weak in-plane electric 
field. In contrast, in 
the regions II, they diffuse and drift to the drains of the areas III. 
Thus, the intensity of the HEP 
should be very sensitive to the exciton diffusion to the regions III 
and a relatively intensive  
HEP can exist only if the diffusion length is comparable or smaller than the stripe width. 

Now, we consider the magnetic-field dependence of the PL intensity. For convenience, the 
spectra at magnetic fields  $0 \ T$ and $7 \ T$ shown in Figs. 4 (a) and (b)  
are normalised to the peak PL intensity for zero voltage difference. Below we consider only 
relative changes in the intensities of the LEP and the HEP  \cite{remark}.
   
One can see from Figs. 4 a, b that the effect of the magnetic field on the normalised intensity of 
the HEP is drastic. 
 We can expect that the increase of the normalised intensity 
of the HEP in a magnetic field is 
associated with a magnetic-field-induced decrease of the exciton diffusion length, 
$l_D=\sqrt{\tau_0D}$, where $D$ and $\tau_0$ are the diffusion coefficient and the lifetime of 
an exciton, respectively.  It was found by a time-resolved PL study in Ref. \cite{Aksenov} that 
the exciton lifetime in QW's decreases approximately by a factor 2 with increasing magnetic 
field from $0 \ T$ to $6 \ T$. The decrease of the lifetime is explained by an increase of the 
radiative recombination probability of an exciton.  In addition, the diffusion coefficient can be 
reduced because of the magnetic-field-induced increase of the exciton effective mass 
\cite{Lerner}. 

In the case of the LEP, the magnetic field does not modify the normalised 
peak PL intensity significantly (see Fig. 5). Two counteracting processes may explain this 
observation. First, the reduction of the 
diffusion length can lead to a decrease of the LEP-intensity, because the LEP arises, in part, 
from excitons diffusing from regions II and III. Second, the magnetic field increases 
the exciton binding energy and, hence, can stabilise excitons against ionisation in 
a strong  in-plane electric field. 
These two mechanisms change the LEP-intensity in opposite ways and, as a result, the PL 
intensity can be relatively 
insensitive to the magnetic field.  
We will see in the next sections, that the magnetic-field
dependence of the PL intensity in lateral superlattices with shorter periods, 
where the 
exciton diffusion length 
is much larger then the stripe width, is different to the above.

\vskip 0.5cm
{\large 4. Model Calculations of the exciton distribution}
\vskip 0.4cm

To support our qualitative explanation of the experimental data, 
we model the transport of two-dimensional 
excitons in a system with lateral and vertical electric fields 
induced by an interdigitated gate. 
The exciton current can be written as  $j_s=-D\nabla n_s+\mu Fn_s$, 
where $n_s$ is the 
surface exciton density, $F=-\nabla U_{eff}$ is the force acting 
on excitons, $U_{eff}(x)$ is 
the 
effective exciton potential as a function of the lateral coordinate  $x$ 
perpendicular to the metal stripes (see 
Fig. 2), and $\mu$ is the mobility of an exciton.  The spatial distribution of 
the exciton density in a sample can be found from the continuity equation, 
$\nabla j_s=-n_s/\tau_0-n_s/\tau_{tun}$, where $\tau_0$ is the exciton 
lifetime in the absence 
of electric fields and $\tau_{tun}$ is the ionisation lifetime of 
an exciton due to an in-plane 
electric field. Thus, the exciton density is determined by the equation 
\begin{eqnarray}
-D\frac{d^2n_s}{dx^2}+\mu\frac{dFn_s}{dx}+\frac{n_s}{\tau_{tot}}=I , 
\label{Diffusion}
\end{eqnarray}
where $I$ describes the generation of excitons by incident light, and 
$1/\tau_{tot}=1/\tau_{0}+1/\tau_{tun}$. 

In our model, the effective exciton potential $U_{eff}$ and the force $F$ 
both depend on the 
vertical 
electric field, while the ionisation lifetime $\tau_{tun}$ depends 
on the in-plane electric field. 
All quantities are periodic functions of the in-plane coordinate 
$x$. To calculate the exciton 
density 
distribution in a sample, we will use a simplified model illustrated in Fig. 6 a.  
We assume 
that the force $F$ is zero in the spatial regions I and II and constant, 
$\pm F_0$, in regions 
III. The strength of the in-plane electric field in the sample depends 
on the geometrical parameters of the  
system and, in particular, on the lateral distance $d=a-w$ between 
the metal stripes. Our model 
calculations of the electric field in the sample show that the maximal 
in-plane electric field 
 in the middle of the regions III is less than $(V_1-V_2)/d$ approximately by a 
factor 2, because the quantum well is separated from the sample surface 
by a distance of $60 \ 
nm$. For our model calculations we choose the gate parameters 
$(a,w)=(1500 \ nm, 1300 
\ nm)$, and the in-plane electric field in the regions III in the form of 
$E_{\parallel}=\pm(V_1-V_2)/2d$. The lifetime $\tau_0$ and 
the diffusion length $l_D$ are 
assumed to be $200 \ ps$ and $700 \ nm$, respectively.  
Analytical expressions for the 
ionisation lifetime of an exciton are known only for 
the strictly 2D and 3D cases in an uniform 
electric field and at zero magnetic field. Here we will 
approximate $\tau_{tun}$ by the 2D 
formula \cite{Miller}, $1/\tau_{tun}\approx36R_y\sqrt{R_y/(eE_{\parallel}a_0)}
\exp{[-32R_y/(3eE_{\parallel}a_0)]}$, where $R_y=4.2 \ meV$ and 
$a_0=14$ $nm$ are the 
Rydberg energy and the Bohr radius in GaAs, respectively. 
In a classical quasi-equilibrium gas 
of excitons, the diffusion coefficient and the mobility are 
connected by the Einstein relation, 
$D/\mu=RT^*$, where $T^*$ is the temperature. 
The temperature of the exciton gas $T^*$ 
can be higher than that of the cryostat bath because of 
the photo-excitation and, for our model 
calculations, is chosen to be $T^*=5 \ K$.  

In our simplified model we can easily solve equation (\ref{Diffusion}), 
assuming that the 
quantities $n_s$  and $j_s$ are continuous functions. 
The calculated exciton density is shown 
in Fig. 6 b.  In fact, the model calculations reproduce 
our qualitative picture described in Sec. 3. 
One can see that the effective force $F=-\nabla U_{eff}$ 
acting on the excitons creates a gradient of the exciton density in the sample. The number of excitons in regions I is increased due to 
exciton flow from regions II.
At the same time, the application of a voltage difference 
results in the appearance of an in-plane electric field in regions III. 
If the value $V_1-V_2$ is small, so that the in-plane field is not sufficient for exciton ionisation, the total number of excitons is conserved.
At high voltage differences the total number of excitons is strongly 
reduced, indicating strong ionisation in the regions III. 

We see that at very high voltage differences the exciton distribution 
is almost symmetric and 
the exciton density in the regions III vanishes because of ionisation. 
The density distribution in 
this case is determined only by the diffusion in the regions I and II, 
where the in-plane electric 
field is absent.  In addition, one can see that the direction of the 
diffusion current at the edges of 
the regions I is changed if the voltage difference exceeds some value. 

The PL spectrum can be found by integrating the function $n_s(x)$, 
\begin{eqnarray}
I_{PL} \propto 
\int dx\frac{\Gamma}{(\omega-\omega_{exc}(x))^2+\Gamma^2}n_s(x) , 
\label{PL}
\end{eqnarray}
where $\omega_{exc}(x)$ is the exciton energy as a function of the coordinate, 
$\Gamma=\Gamma_0+1/\tau_{tun}$, and $\Gamma_0$ is the half-width 
of the PL exciton 
peak in the absence of a potential modulation. 
From the data in Fig. 3 we find that, 
in the vicinity of the voltage $-0.9 \ V$, the 
exciton energy in our QW at zero magnetic field can be approximated by 
$\omega_{exc}(x)=1515 meV+\alpha[E_z(x)-E_0]/E_0$, where $E_z(x)$ 
is the vertical electric field induced by the applied voltage, $E_0=2.3\times10^4 \ V/cm$, 
and $\alpha=16 \ meV$.  
Figure 7 shows the calculated PL spectra at various voltage differences. 
At finite voltage differences the PL structure is split into two peaks of different intensities. 
The LEP is more intense than the HEP because of drift and diffusion of excitons from the  
regions II to the regions I. At high voltage differences the peaks become almost symmetric. 

We can see from our simulations that the intensity of the HEP is more sensitive 
to the diffusion coefficient than that of the LEP. This fact is connected with the behavior of 
the exciton density near the edges of the regions I and II.
Near the edges of the regions II, 
the exciton density is reduced because of drift to the regions III.
At the same time, this factor leads 
to an increase of the exciton density near the 
edges of the regions I as long as the excitons are not ionised.  
Mathematically, if the exciton 
density at the edges of the regions II becomes small, 
the exciton distribution in the regions II is 
determined only by the diffusion equation with the 
boundary conditions $n_s=0$ at the edges.  

Using this fact, we can roughly estimate the diffusion 
length by use of our experimental data 
for the HEP intensity at a high voltage difference. 
If $n_s=0$ at the edges of the regions II, 
the exciton density distribution in these regions is 
$n_s=I\tau_0[1-\cosh(x/l_D)/\cosh(w/2l_D)]$. 
The relationship between the intensity of the HEP, $I_{HEP}$, 
and the intensity of the PL 
exciton peak at zero voltage difference, 
$I_{0}$, is $I_{HEP}/I_0=(w/2a)[1-(2l_D/w)\tanh(w/2l_D)]$. 
To estimate the diffusion length we use the data presented in Fig. 8. 
In this figure, the 
normalised PL spectrum is shown for the case when the voltage 
$V_2$ is kept constant, while 
the voltage $V_1$ is varied.  The perpendicular electric field 
beneath the gate 2 is kept constant 
and the HEP intensity does not change at high voltage 
differences (see Fig. 8). The HEP at 
$B=0 
\ T$ in Fig. 8 is hardly visible so that we give here 
estimations for finite magnetic fields.  From 
the data shown in Fig. 8 at high voltage differences 
we have approximately: 
$I_{HEP}/I_0\approx0.07$ and  $0.13$ for magnetic fields $B=3.5 \ T$ and $7 \ T$, 
respectively. These data correspond to diffusion 
lengths of $800 \ nm$ and 
$500 \ nm$ for the cases $B=3.5 \ T$ and $7 \ T$, respectively. 
The clear decrease of the 
diffusion length 
with increasing magnetic field 
can be connected both with the reduced exciton 
lifetime $\tau_0$ 
and the reduced diffusion coefficient $D$ 
as was discussed in Sec. 3.  
We note that our estimate of the diffusion 
length is in reasonable agreement with the data of 
spatially resolved studies of the exciton 
diffusion in quantum wells \cite{Heller}. 

Using our model we can qualitatively describe 
the experimentally observed behavior of the PL 
spectra at various voltages shown in Figs. 4 a, b. 
We note, however, that even at a high voltage difference $\Delta V$ 
the intensity of the LEP remains larger than that of the HEP, 
while in our model the peaks at a high $\Delta V$ become symmetric. 
This disagreement between our model and the 
experimental data may be connected with the behavior of 
the effective exciton potential near the edges of the metal stripes.
In the experimental situation the effective exciton potential near 
the edges of the stripes is smoother than our model potential. 
This will reduce the exciton current to the drain regions and, hence, increase the LEP intensity.

\vskip 0.5cm
{\large 5. Photoluminescence in short-period lateral superlattices}
\vskip 0.4cm
   
The intensity of the HEP is a function of the 
exciton diffusion lengths and the lateral sizes of the structure.  
We can expect that the ratio of the intensities of HEP and LEP, 
$I_{HEP}/I_{LEP}$, decreases with  decreasing period of 
the lateral superlattice.  In other 
words, in short-period superlattices almost all excitons 
can reach the potential trap under the 
gate 1 and, thus, the HEP will be less intense. 
Corresponding behavior of the experimental spectra is demonstrated  
in Fig. 9.  For convenience, the spectra of Fig. 9 are normalised to the 
peak intensity of the LEP. The 
spectra are recorded in a strong magnetic field, where the  
HEP becomes more pronounced.  
They show a clear decrease of the ratio $I_{HEP}/I_{LEP}$ 
with decreasing gate period. 
We note that the HEP in the sample with the smallest superlattice period 
$2a=500 \ nm$ at zero magnetic field is almost invisible. 
   
Another interesting feature of the sample with the smallest 
superlattice period of $500 \ nm$ is the voltage-dependence of the LEP intensity. 
Figure 10 shows  the PL intensity of the LEP as a function of 
the voltage $V_2$ when the voltage $V_1$ is kept constant.  
The LEP energy depends on the QCSE in the regions I and is hardly changed 
with the voltage $V_2$.  
However, the voltage $V_2$ has a strong influence on 
the PL intensity.
The intensity in Fig. 10 decreases in two steps. The first step occurs at 
$V_2 \approx-0.1 \ V$.
We explain the first step in the voltage-dependence of 
the intensity by the ionisation of excitons 
moving from the regions II to the regions I due to a strong 
in-plane electric field in the areas III between 
the metal stripes.  In this process almost 50\% of the excitons in the 
sample are destroyed.  This decrease of the intensity is similar to 
that described in the section 2 for the case of 
the long-period superlattice.  At the voltage $V_2 \approx0.15 \ V$ the voltage-dependence of 
the intensity exhibits a second step. When the 
voltage 
$V_2$ is larger than $0.2 \ V$ the PL intensity vanishes. 
The second step is not observed in 
long-period superlattices (see e. g. Fig. 5) and 
arises from exciton ionisation in the regions 
I. We think  that in short-period lateral superlattices 
such an ionisation process is possible even 
in the centers of the regions 
I. Here  the electrostatic potential is nearly parabolic (see Fig. 2 c). 
In the center of the regions I, the hole bound in the 
exciton is strongly localised but the corresponding electron can 
tunnel if the curvature of the parabolic electron 
potential 
is sufficiently strong.  In the system with the interdigitated 
gate, the characteristic energy of the 
parabolic potential in the middle of the metal stripe can be about $10 \ 
meV$ \cite{Drexler94}.  
This value 
is comparable to the exciton binding energy in a $20$-$nm$-QW. 
Thus, in small period superlattices where the obtained electron potential modulation is essentially parabolic, 
ionisation becomes possible even in the center of the regions I. 
We 
note that the data of Fig. 10 are recorded for 
an unoccupied QW. 
   
The voltage-dependence of the PL intensity at small voltage differences 
for $B=7\ T$ 
in Fig. 10 shows an 
unexpected non-monotonic behavior. 
 We suppose that the slight increase of the PL at small voltage 
differences in Fig. 10 can arise from a change of 
the vertical electric field in a sample due to a 
redistribution of surface charges with application of the voltage difference. 
    
We observe that the ionising voltage differences responsible 
for the two-step behavior are 
increased by the application of the magnetic field (see Fig. 10). 
In figure 11 we plot the normalised peak PL intensity, 
i. e. the value 
$I(\Delta V)/I(\Delta V=0)$.   Because the ratio 
$I(\Delta V)/I(\Delta V=0)$ for $\Delta V \neq0$ 
increases with magnetic field, the PL intensity 
in a lateral superlattice increases faster than 
in a homogeneous system.  We attribute this 
behavior to the stabilisation of excitons by a 
high magnetic field. The latter can be connected 
with a suppression of electron tunnelling and an 
increase of the exciton binding energy in magnetic field.

\vskip 0.5cm
{\large 6. Summary}
\vskip 0.4cm

We have studied the PL spectra of a QW system with interdigitated top gates of various 
periods.  The PL spectra clearly demonstrate the localisation of two-dimensional excitons in 
linear traps induced by the QCSE.  It is shown that the QCSE can be used as an effective 
mechanism to induce excitonic traps in laterally microstructured electro-optic devices.  The 
behavior of the PL intensity at various voltages are interpreted with a model involving an 
effective exciton potential. By use of magneto-PL studies we observe two regimes of exciton 
transport in our samples. 
The first is realised at small voltages, when the excitons, created in all lateral regions of the 
sample, can diffuse and drift to regions with minimal effective exciton potential. This regime 
becomes possible because the exciton diffusion length in our samples 
is comparable with the lateral gate periods.  
The second regime of exciton transport 
occurs at relatively high voltage differences, when an 
in-plane electric field between the finger gates is sufficient to ionise excitons. In this regime, 
the regions under the gates 1 and 2 are not coupled by an exciton current. 
The magnetic field has a strong influence on the PL spectra, which is 
discussed in terms of  suppression of exciton diffusion and exciton stabilisation.

We wish to note that localisation of 2D excitons in linear traps via the QCSE can also be 
obtained by means of a 
surface acoustic wave (SAW) which indudes a strong modulation of the vertical electric field in a 
sample with a QW. In such a system, the excitonic traps are moving with the sound velocity 
of GaAs, $c_s=3*10^5\ cm/s$.  With a typical exciton lifetime of $300 \ ps$, the excitons 
localised in the effective potential minima induced by the SAW can be transferred over 
typical distances  
of $1 \ \mu m$, comparable to  characteristic lateral sizes 
of semiconductor microdevices. Thus, a 
combination of SAW and the QCSE can be used to transfer optically-active excitons between 
different elements of a microdevice.  At present, such acousto-optics of QW«s 
is a subject of active experimental investigations \cite{Rocke96}.

\vskip 0.5cm
{\large Acknowledgements}
\vskip 0.4cm

This work was supported by the Sonderforschungsbereich 348 of the Deutsche 
Forschungsgemeinschaft.  One of us (A. O. G.) gratefully 
acknowledges support by the A. v. Humboldt Stiftung.

\newpage

* permanent address:  Institute  of Semiconductor Physics, 630090 Novosibirsk, Russia.

** permanent address:  Institut f\"ur Angewandte Physik, Universit\"at Hamburg, 
Jungiusstrasse 11, D-20335 Hamburg, Germany.

\newpage
\ \\
{\Large{\bf Figure captions}}
\vskip 1.cm
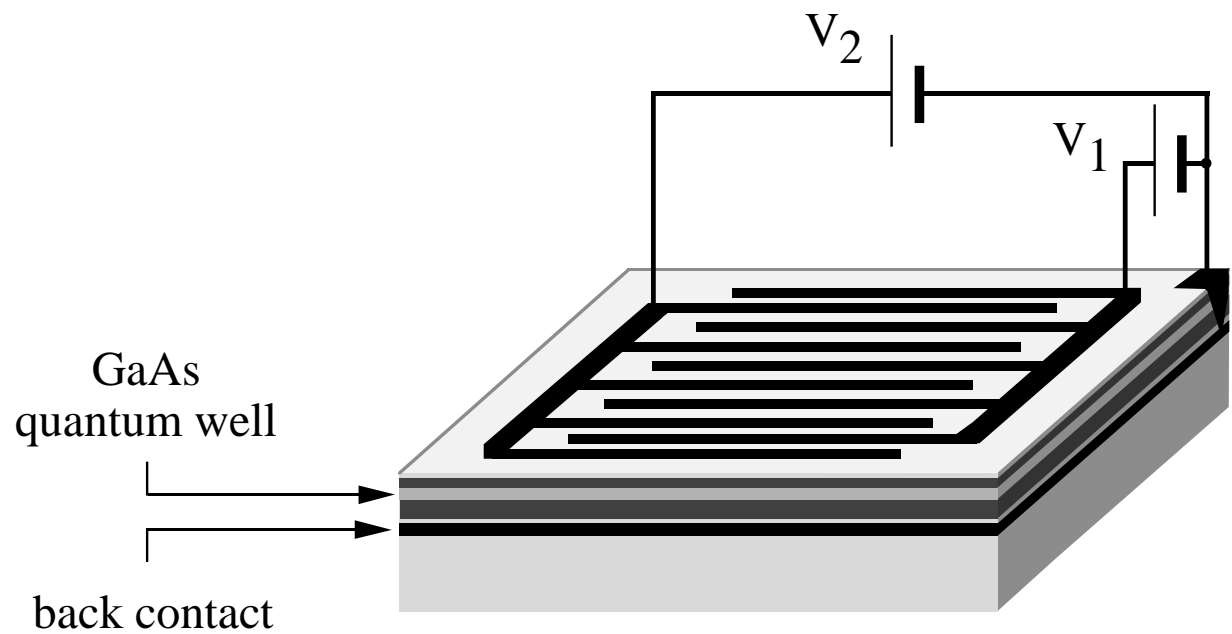
\ \\ {\bf Fig. 1.} \\
Sketch of a QW system with an interdigitated metal top gate. The voltages $V_1$ and 
$V_2$ are applied to the finger gates with respect to the back contact. 
   
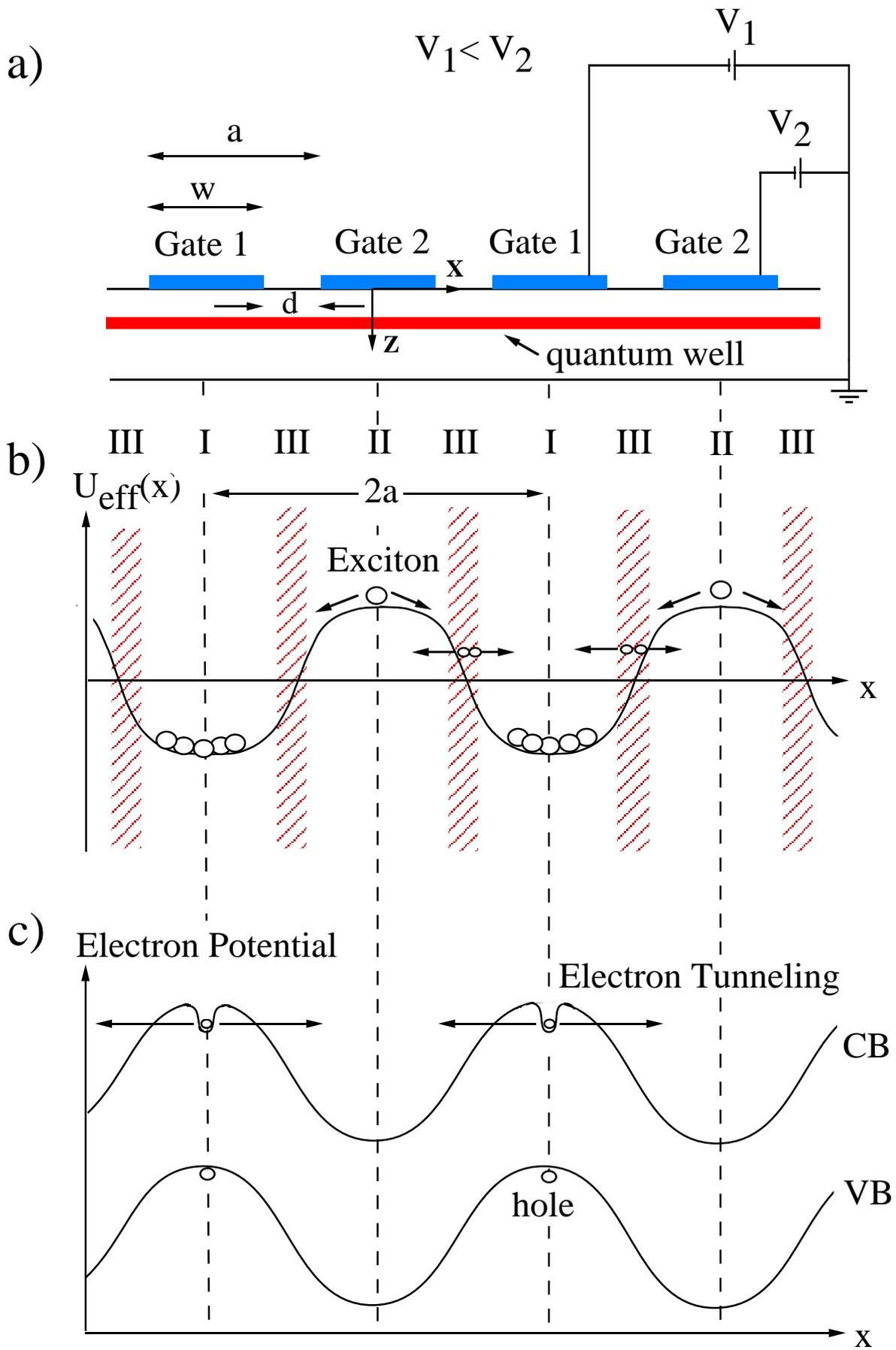
\ \\ {\bf Fig. 2.} 
\ \\  a) The geometry of the system with an interdigitated gate. 
\ \\  b) The effective exciton potential, $U_{eff}$, induced 
by the QCSE as a function of the 
in-plane coordinate $x$. The regions I and II are located 
under the gates 1 and 2, respectively. The 
dashed regions III correspond to the areas of  
strongest in-plane electric fields. Also, we 
sketch schematically the picture of exciton 
transport in the sample. 
\ \\  c) The electron potential of our system at 
a finite voltage difference $\Delta V$ applied to the 
interdigitated gate. The excitons are localised 
in the regions I, where the electron potential is 
maximal and nearly parabolic. The dip in the CB is caused by the Coulomb interaction of the electron with the hole.
   
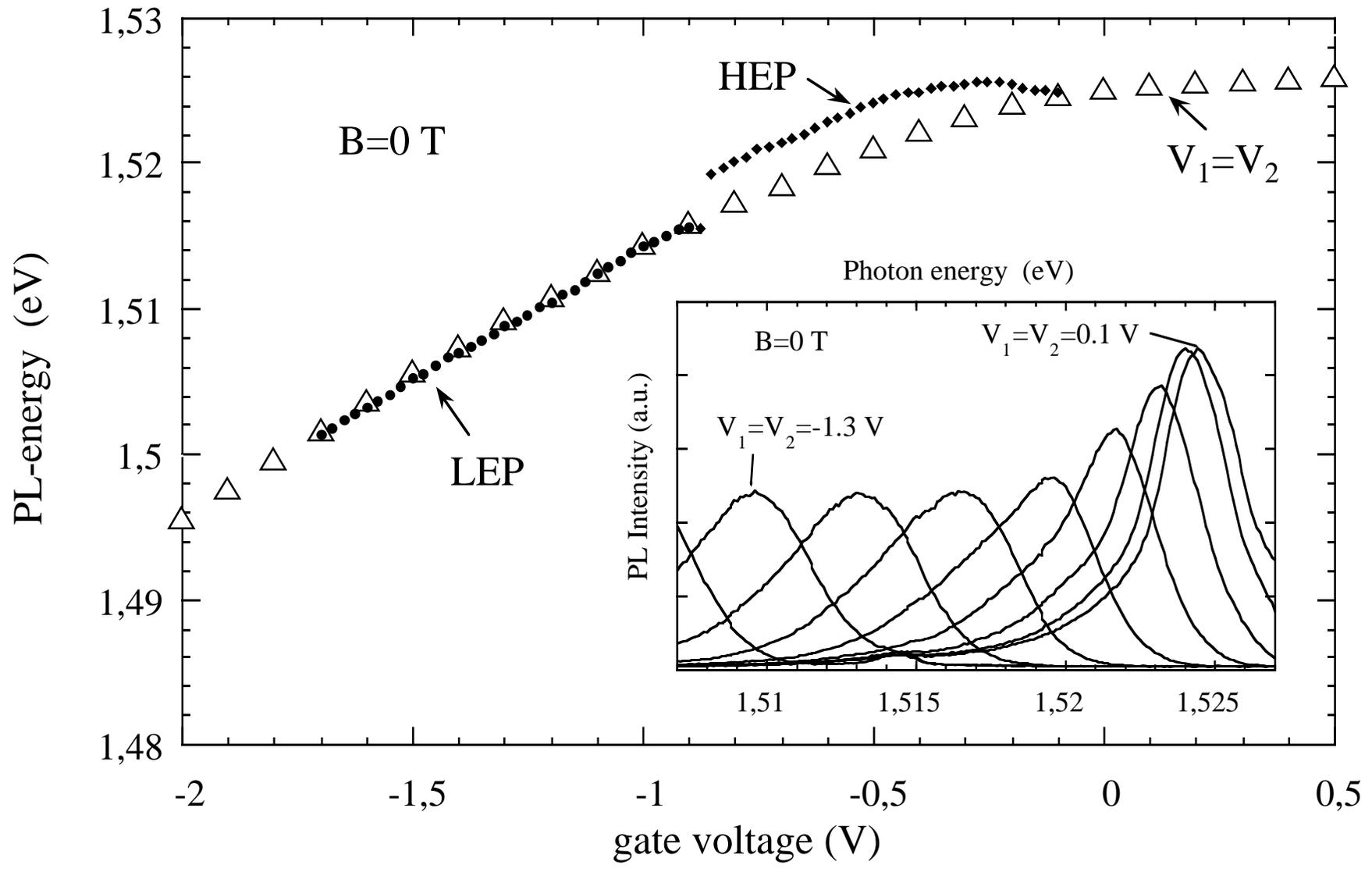
\ \\ {\bf Fig. 3.} \\
Energy  positions of the exciton peak as a function of equal applied voltages $V_1=V_2$ at 
zero magnetic field (open symbols). The corresponding spectra are displayed in the inset.  
For comparison the energy positions of the low-energy peak (LEP) and the 
high-energy peak (HEP) in the sample with the lateral superlattice 
of period $2a=3000 \ nm$
are depicted 
by filled circles and rectangles, respectively. Here the average voltage $(V_1+V_2)/2$ is kept 
constant at $-0.9 \ V$.  
The energy of the low-energy peak is plotted as a function of voltage $V_1$, which is 
varied from $-0.9 \ V$ to $-1.7 \ V$. Similarly, the energy of the high-energy peak is shown as 
a function of voltage $V_2$, which is varied from $-0.9 \ V$ to $-0.1 \ V$.    

\ \\ {\bf Fig. 4.} 
\\   PL spectra of the sample with the $3000$-$nm$-superlattice period (a) 
at zero magnetic field 
and (b) at a magnetic field of $B=7T$ applied perpendicular to the sample surface.
The intensities are normalised to the one at zero voltage difference 
$\Delta V = V_1-V_2$. The spectra are offset for clarity.
The average voltage is $(V_1+V_2)/2=-0.9 \ V$, and 
the voltage difference 
$\Delta V$ is changed from $0 \ V$ (upper curve) to 
$-1.8 \ V$ (lower curve) in steps of $0.1 \ V$
   
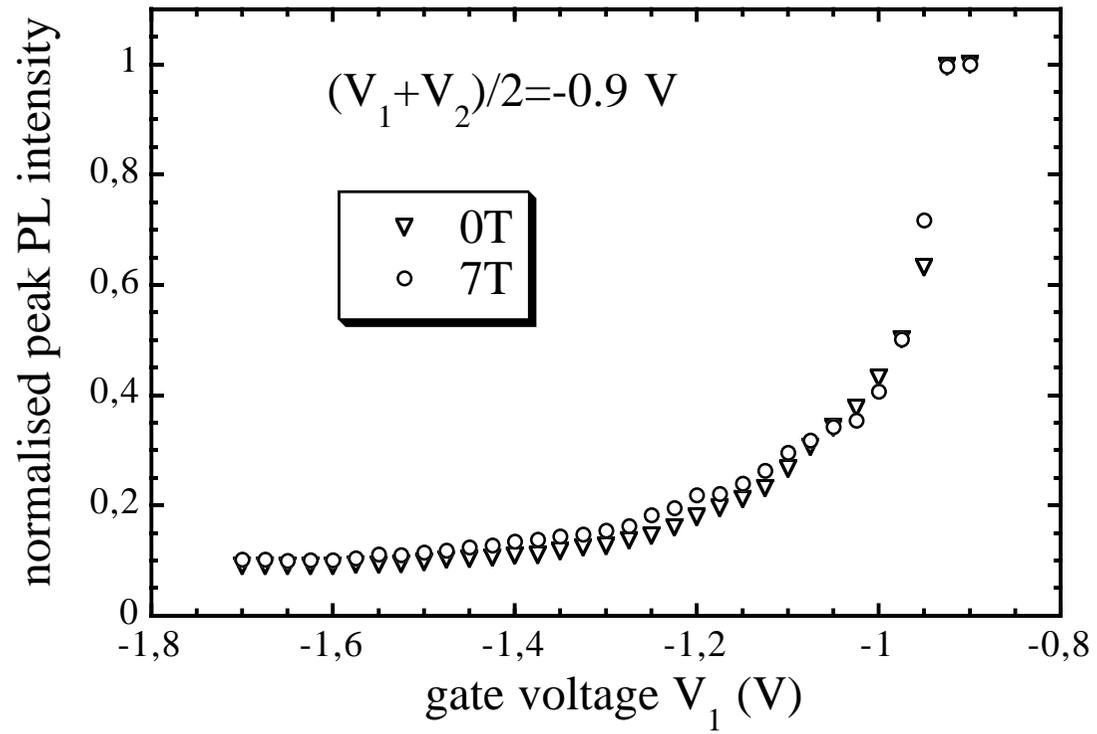
\ \\ {\bf Fig. 5.} \\
PL intensity of the low-energy peak as a function of the voltage 
$V_1$ for magnetic 
fields $0 \ T$ and $7 \ T$. The voltages 
$V_1$ and $V_2$ are set as described in Fig. 4. The 
intensity is normalised to the one for zero voltage difference 
$\Delta V=V_1-V_2$.  
   
\ \\ {\bf Fig. 6.} \\ 
a) A model for a description of the exciton transport in a sample: the solid and dashed lines 
show schematically the behavior of the functions $U_{eff}$ and $1/\tau_{tun}$, respectively.
\\ b) The calculated spatial distribution of the exciton density at various voltage differences; the 
characteristic density is $n_0=I\tau_0$. 
  
\ \\ {\bf Fig. 7.} \\ 
Calculated PL spectra at various voltage differences $\Delta V=V_1-V_2$; the average voltage is 
kept constant at 
$(V_1+V_2)/2=-0.9 \ V$ and the broadening parameter is set to $\Gamma_0=0.7 \ meV$. The parameters $l_D$, $T^*$, and $\tau_0 $ are explained in the text. 

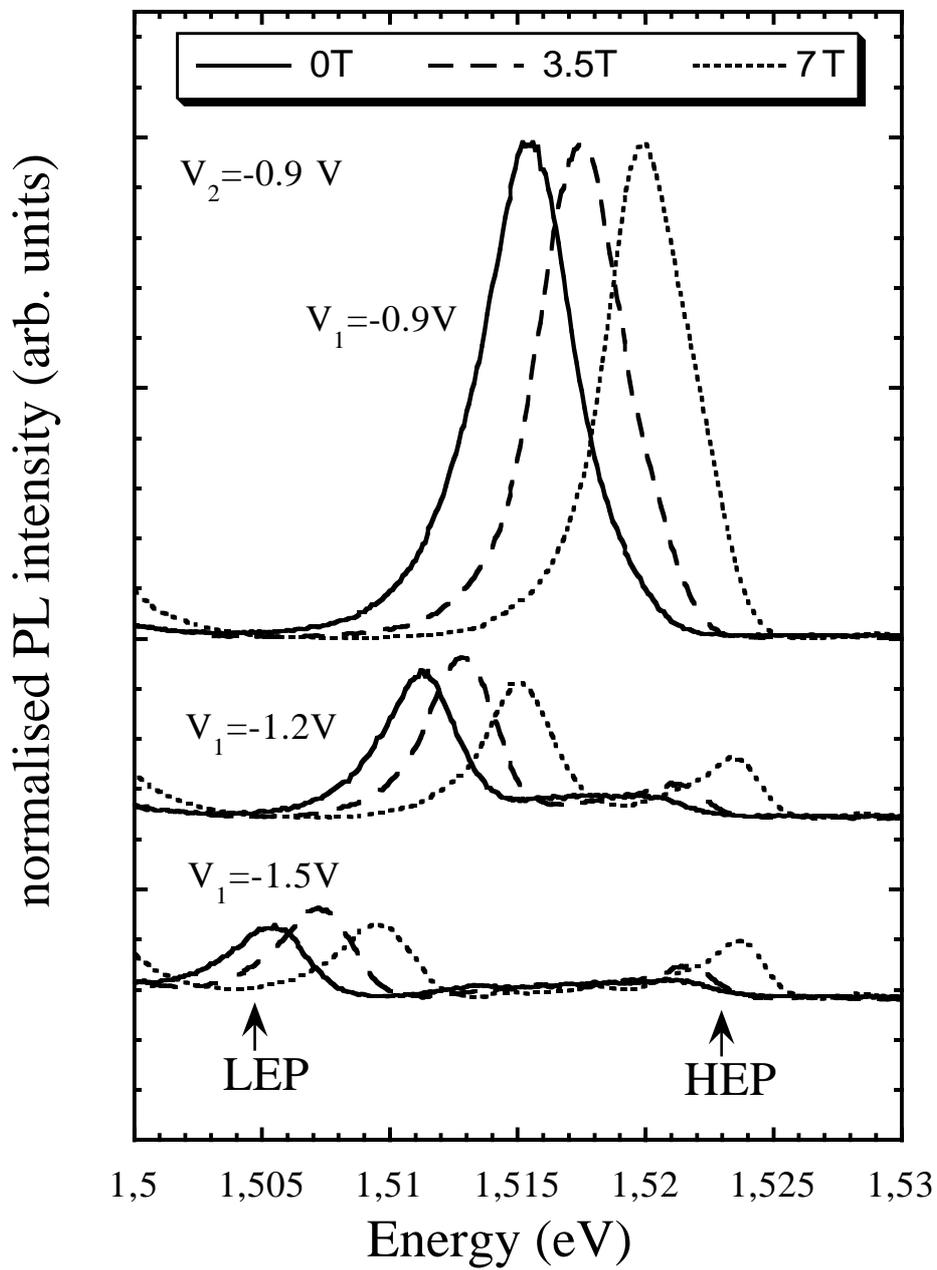
\ \\ {\bf Fig. 8.} \\ 
Experimental data recorded at various voltage differences $\Delta V=V_1-V_2$ 
and magnetic fields for the lateral superlattice 
of period $2a=3000 \ nm$. 
The voltage $V_2$ is kept constant, while the voltage 
$V_1$ is changed. The spectra are offset for clarity.
  
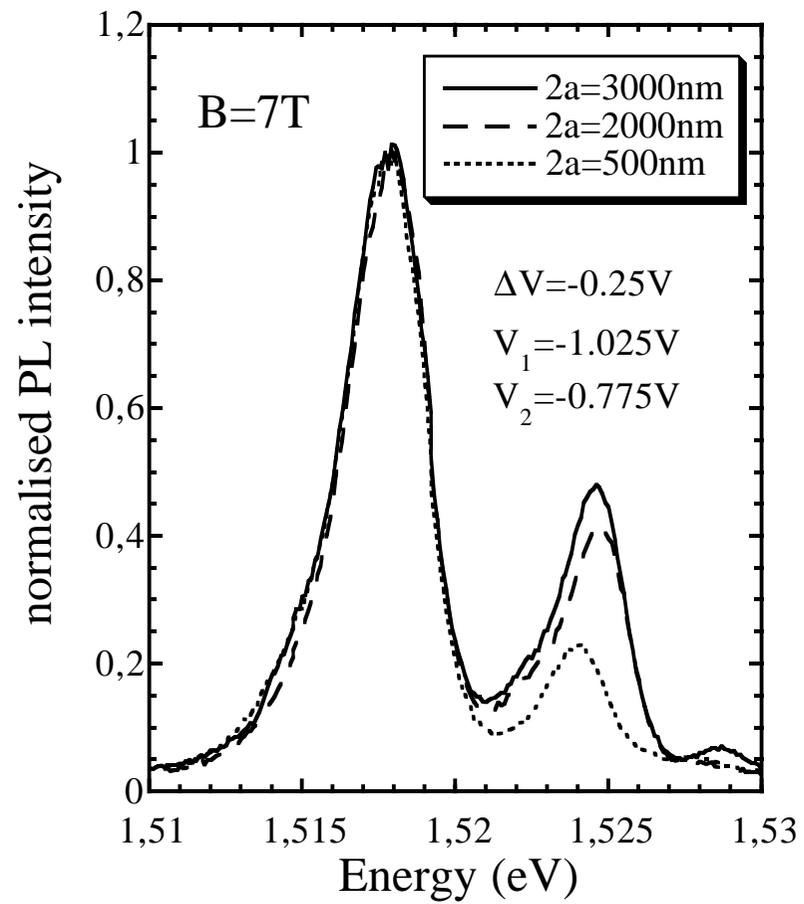
\ \\ {\bf Fig. 9.} \\
PL spectra of samples with three different periods at magnetic field $B=7 \ T$ and 
applied voltages $V_1=-1.035 \ V$ and $V_1=-0.775 \ V$. The spectra are normalised to the 
intensity of the low-energy peak. 

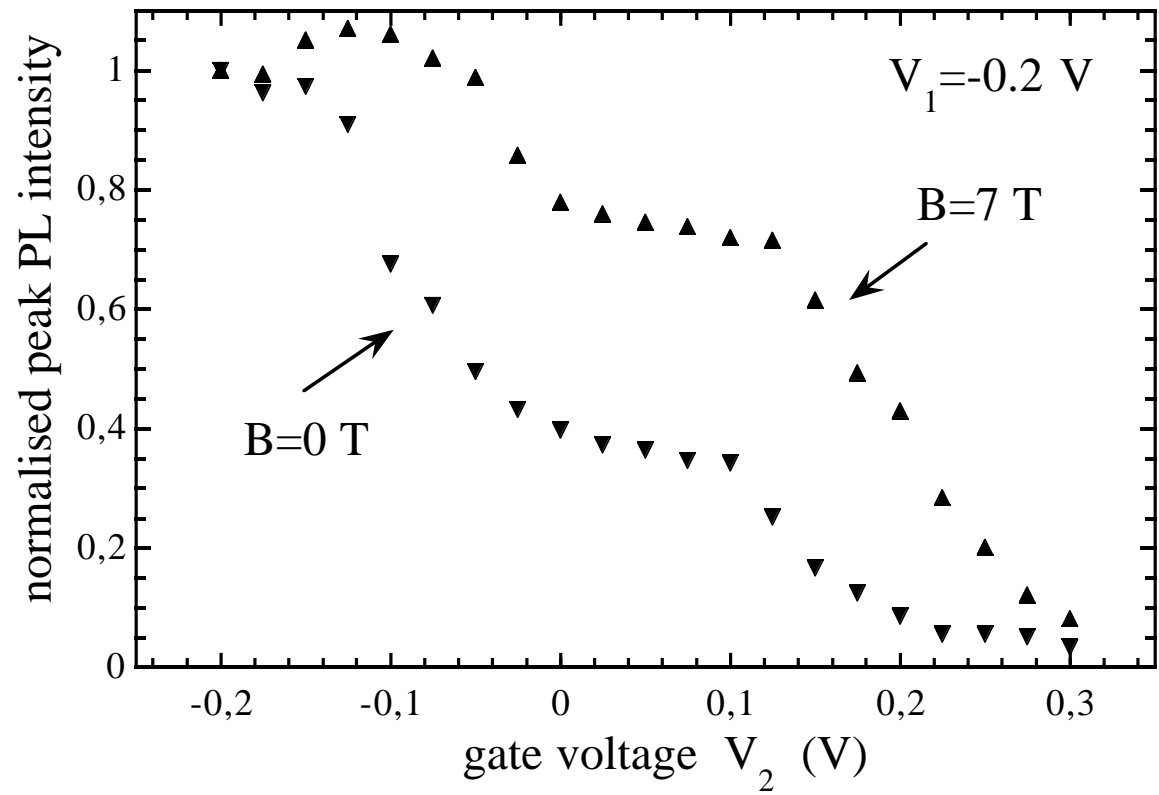
\ \\ {\bf Fig. 10.} \\ 
The peak intensity of the low-energy peak in  the lateral 
superlattice with period $2a=500$ $nm$ 
as a function 
of the voltage $V_2$ for magnetic fields $0 \ T$ and $7 \ T$. 
The voltage $V_1$ is kept constant at $-0.2 \ V$. 
   
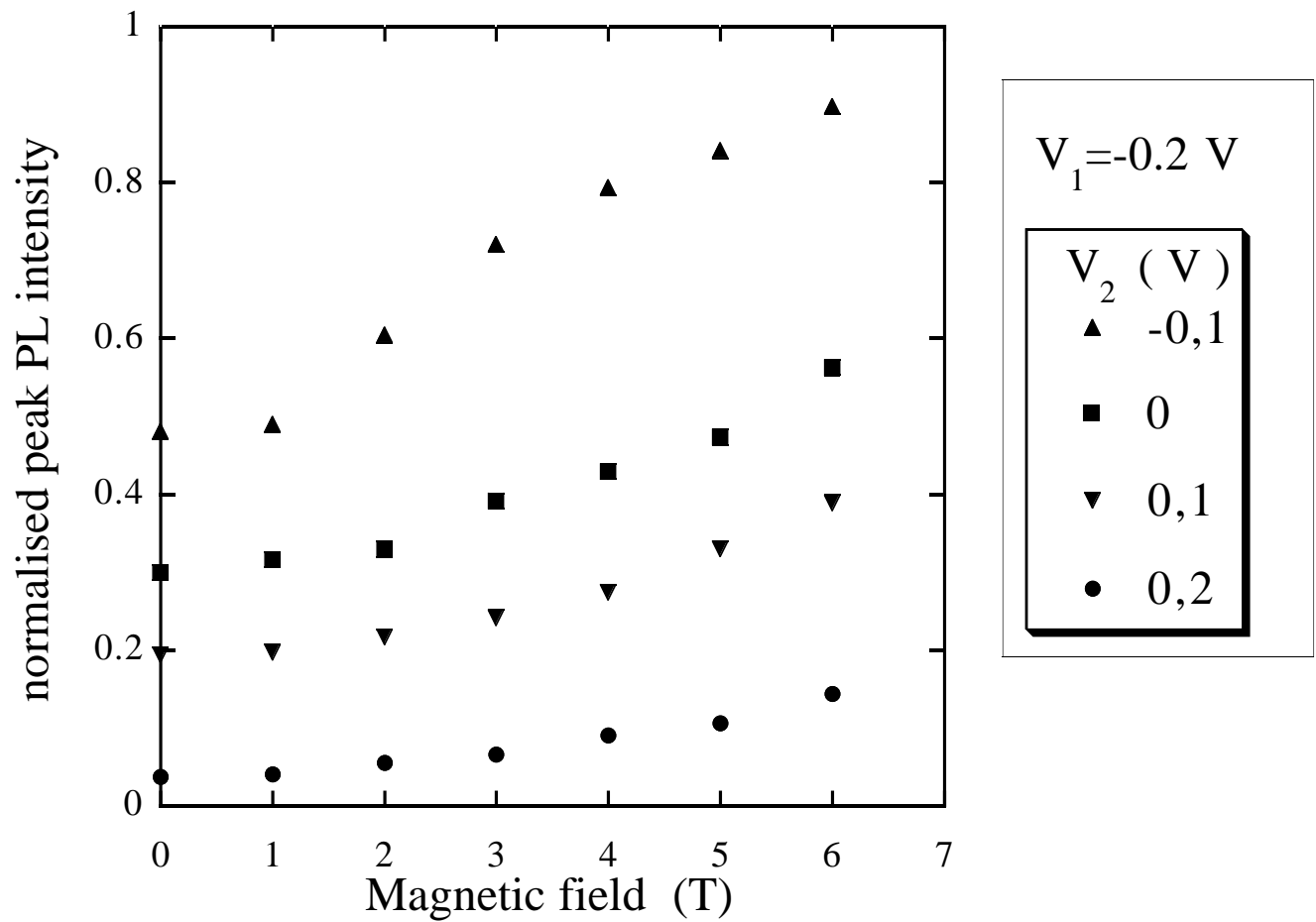
\ \\ {\bf Fig. 11.} \\ 
The peak intensity of the low-energy peak in 
the lateral 
superlattice with period $2a=500$ $nm$ 
as a function 
of the magnetic field for a few voltage differences. The peak intensity is 
normalised to that for zero voltage difference.


\vskip 1.cm

\begin{thebibliography}{99}

\bibitem{Schmitt-Rink} 
For a review, see e. g., 
{\it Optics of Semiconductor Nanostructures}, ed. by F. Henneberger, 
S. Schmitt-Rink, and E. O. G\"obel  (Akademie Verlag, Berlin, 1993). 

\bibitem{Drexler94} 
H. Drexler, W. Hansen, S. Manus, J. P. Kotthaus, M. Holland, and S. P. Beaumont, Phys. 
Rev. {\bf B49}, 14074 (1994);  H. Drexler, W. Hansen, J. P. Kotthaus, M. Holland, and S. 
P. Beaumont,  Appl.  Phys. Lett.  {\bf B64}, 2270 (1994).

\bibitem{Kash} 
K. Kash, J. M. Worlock, M. D. Sturge, P. Grabbe, J. P. Harbison, A. Scherer, and 
P. S. D.  
Lin, Appl. Phys. Lett. {\bf 53}, 782 (1988); Y. Gu, M. D. Sturge, K. Kash,  N. Watkins,  
B. P.  Van der Gaag,  A. S. Gozdz,  L. T. Florez, and J. P. Harbison, $ibid$ {\bf 70},  1733 
(1997).  

\bibitem{Brunner} 
K. Brunner, G. Abstreiter, M. Walther, G. B\"ohm, and G. Tr\"ankle,  Surf. Sci. {\bf 267}, 
218 (1992). 

\bibitem{Marzin} 
J. Y. Marzin,  J. M.  Gerard, A. Izrael, D. Barrier, and G. Bastard,  Phys. Rev. Lett. {\bf 73}, 
716 (1994).

\bibitem{Lederman} 
F. L. Lederman and J. D. Dow, Phys. Rev. {\bf 13}, 1633 (1976). 


\bibitem{exp} 
Here we wish to note that the Stark red shift caused by the in-plane electric field was not visible in recent optical experiments on quantum wells (see Refs. 12, 13). 

\bibitem{Miller} 
D. A. B. Miller, D. S. Chemla, T. C. Damen, A. C. Gossard, W. Wiegmann, T. H. Wood, 
and C. A. Burrus,  Phys. Rev. {\bf B32}, 1043 (1985).

\bibitem{Vina} 
L. Vina, E. E. Mendez, W. I. Wang, L. L. Chang, and L. Esaki, J. Phys. C: Solid 
State Phys. {\bf 20}, 2803 (1987).

\bibitem{Schmeller95} 
A. Schmeller, A. Govorov, W. Hansen, J.P. Kotthaus, W. Klein, G. B\"ohm, G. Tr\"ankle, 
and G. Weimann,  Proceed. {\it 22nd Int. Conf. on the Physics of Semiconductors}, Canada, 
1994 
(World Scientific, 1995), V.2, p.1727.

\bibitem{Markus} 
M. Hagn,  A. Zrenner, G. B\"ohm, and G. Weimann,   Appl. Phys. Lett  {\bf 67}, 232 
(1995). 


\bibitem{Schmeller94} 
A. Schmeller, W. Hansen, J.P. Kotthaus, G. Tr\"ankle, and G. Weimann, Appl. Phys. Lett.  
{\bf 64}, 330 (1994). 

\bibitem{Dohler} 
M. Kneissl, N. Linder, P. Kiesel, S. Quassowski, K. Schmidt, 
G. H. D\"ohler, H. Grothe, and J. S. Smith, 
{\it "Superlattices and Microstructures"}  {\bf 16}, 109 (1994).


\bibitem{remark0} 
Flattening of the electrostatic potential at the equal non zero voltages $V_1=V_2$ 
arises from a redistribution of surface charges between the stripes under the 
interband illumination. 
Nevertheless, there is a small remaining modulation, which will be discussed below. 

\bibitem{Heller} 
W. Heller, A. Filoramo, Ph. Roussignol, 
and U. Bockelmann, Solid-State Electr. {\bf 40}, 725 
(1996). 

\bibitem{remark} 
We cannot absolutely compare the PL intensities at different 
magnetic fields because varying the magnetic field changes slightly 
the position of the sample rod and thus results in a small shift of 
the sample with respect to focus of the collection optics. 

  
\bibitem{Aksenov} 
I. Aksenov, Y. Aoyagi, J. Kusano, T. Sugano, T. Yasudo, and Y. Segawa, Phys. Rev. {\bf 
B52}, 17 430 (1995).

\bibitem{Lerner} 
I. V. Lerner and Yu. E. Lozovik, JETP {\bf 51}, 588 (1980). 

\bibitem{Rocke96} 
C. Rocke, S. Zimmermann, A. Wixforth, J. P. Kotthaus, H. B\"ohm, and G. Weimann,  
Phys.  Rev.  Lett., in press. 



\end{thebibliography}
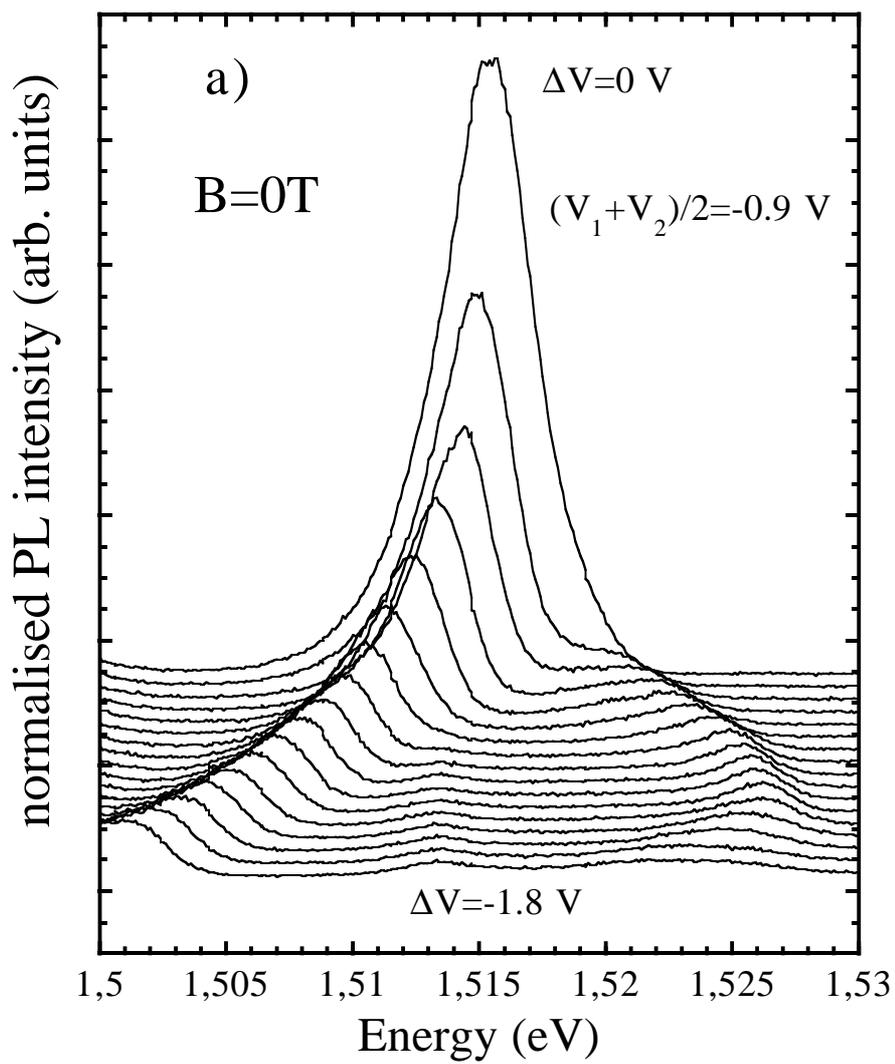
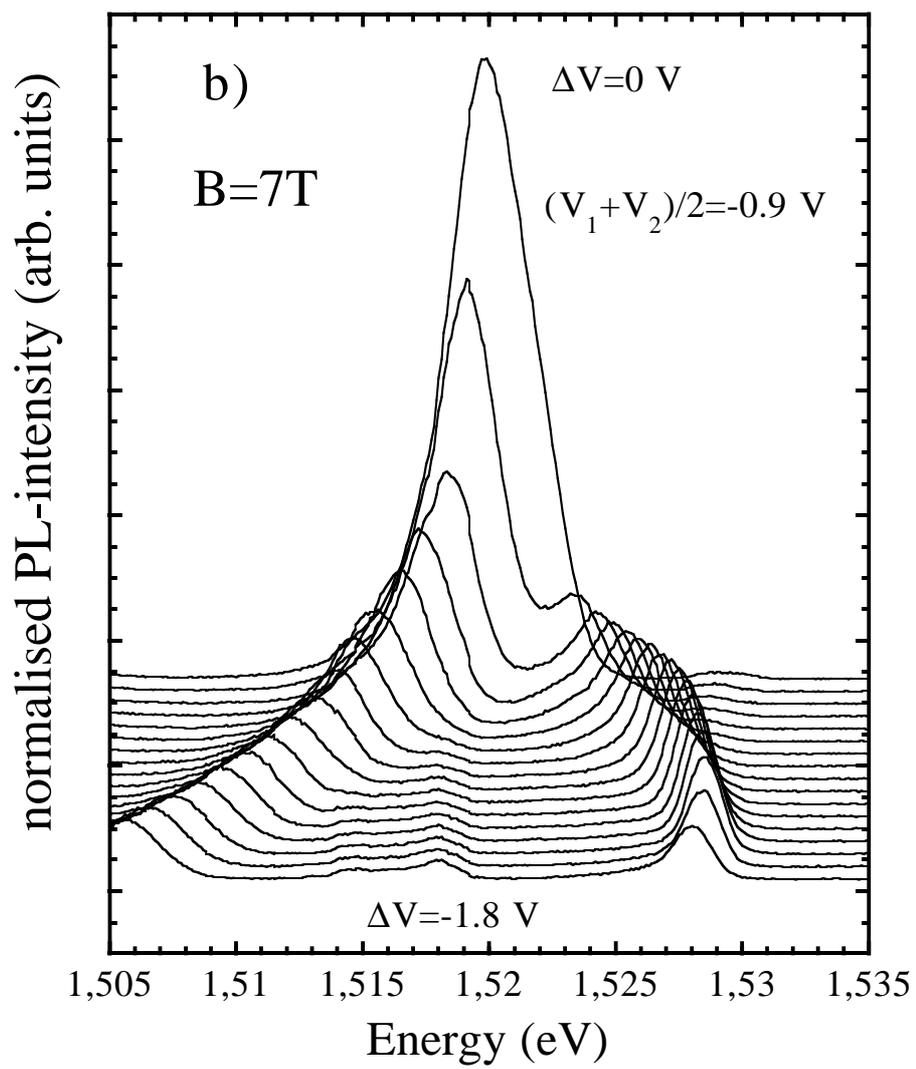
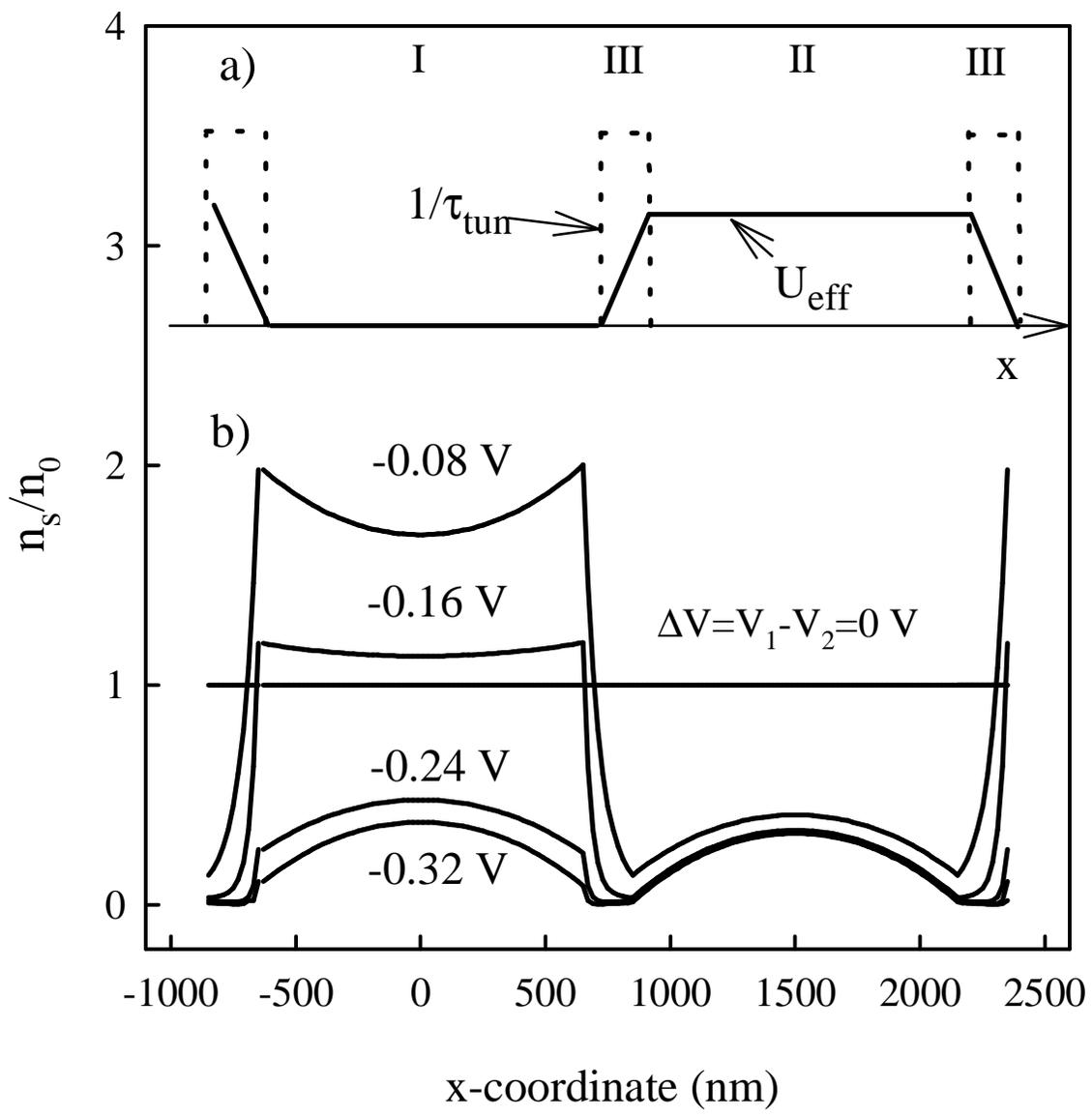
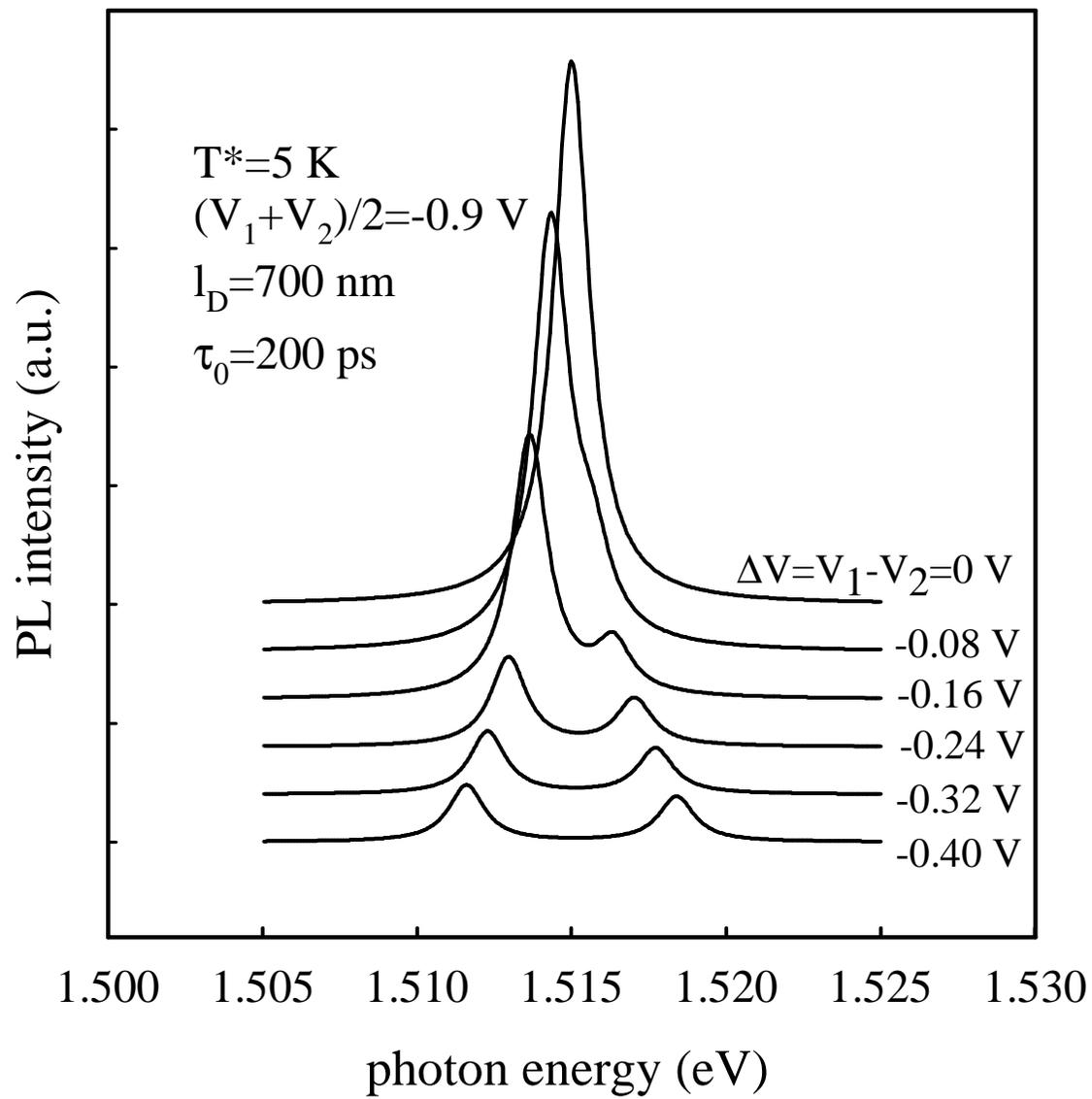
\end{document}